\documentstyle [12pt]{article}

\begin{document}

\vspace{10mm}
\centerline{\Large\bf Locating analytically critical temperature}
\centerline{\Large\bf in some statistical systems }
\vspace*{1cm}
\begin{center}
{\large J. Wosiek}
\\
\vspace*{1.5mm}
Institute of Physics, Jagellonian University \\
Reymonta 4, 30-059 Cracow, Poland
\end{center}
\vspace*{1.5cm}
\hspace*{\fill} \fbox{To my Father} \newline
\vspace*{1.5cm}
\begin{abstract}
We have found a simple criterion which allows for the straightforward
determination of the order-disorder critical temperatures.
The method reproduces exactly  results known for the two dimensional Ising,
Potts and $Z(N<5)$ models. It also works for the Ising model on the
triangular lattice. For systems which are not selfdual our proposition remains
an unproven conjecture.
It predicts $\beta_c=0.2656...$ for the two coupled
layers of Ising spins. Critical temperature of the three dimensional
Ising model is related to the free energy of the two layer Ising system.
\end{abstract}
\vspace*{2cm}
PACS number: 64.60.Cn \newline
hep-lat/9312025
\newline
TPJU-23/93 \newline
November 1993 \newline
\newpage

 Classical method of determining critical temperature in statistical physics
 consists
of locating singularities of the largest eigenvalue of the corresponding
transfer matrix
${\cal T}$. To this end one has to study the high power of ${\cal T}$, e.g.
${\cal T}^L$ with
$L$ being the linear size of the system. This amounts to investigate the $d$
dimensional
euclidean system in its full complexity. On the other hand it is conceivable
that the
information about the phase transition is also encoded in all other
eigenvalues, i.e. that
the whole spectrum of the transfer matrix is sensitive to its location.
This observation is confirmed
by the famous example of the two dimensional Ising model where {\em all}
eigenvalues show the extremal behaviour at $\beta=\beta_c$ \cite{huang}.
Therefore we propose to search for the extremum of some simpler
function which characterizes the system, which however is not
dominated solely by the largest eigenvalue. The advantage of such an
approach is that one may infer a nontrivial information from
"characteristic" functions which are much easier to calculate.
To be specific we propose to study the following characteristic function
of a $d$ dimensional system
\begin{equation}
\rho(\beta) = \lim_{L\rightarrow\infty} \left( {(Tr {\cal T})^2 \over
Tr {\cal T}^2 }
\right )^{{1\over L^{d-1}}} ,  \label{rho}
\end{equation}
and analogous higher, moments of the transfer matrix.
Following simple properties of $\rho$ can be easily proven.
\begin{eqnarray}
a)  & \rho(0)=1, \nonumber \\
b)  & \rho(\infty)=1, \\
c)  &  \rho(\beta) \ge 1 .     \nonumber
\label{abc}
\end{eqnarray}
To see $a)$ consider the normalized second moment at finite $L$
\begin{equation}
r_L(\beta)={(Tr {\cal T})^2 \over Tr {\cal T}^2 }= {(Z_1)^2 \over Z_2},
\label{rl}
\end{equation}
where $Z_1$ and $Z_2$ are the partition functions of the $d-1$ dimensional
system and of the two coupled $d-1$ dimensional
systems respectively. For simplicity we shall use the
terminology of the d=2 spin system. Hence $Z_1$
and $Z_2$ describe one dimensional
spin chain and two coupled chains of spins. Periodic boundary conditions are
implied
in both directions, even for the single chain. With these definitions $a)$
follows immediately from the observation that
$Z_2(\beta=0)={\cal N}_2=({\cal N}_1)^2$, where
${\cal N}_1, {\cal N}_2$ denote the total number of microscopic states of
single and double
chain.  At low temperatures $(\beta \rightarrow \infty)$ fully ordered states
dominate, hence
\begin{equation}
\rho(\beta=\infty)=\lim_{L\rightarrow \infty} g^{1\over L^{d-1}} =1.
\end{equation}
Where $g$ denotes degeneracy factor of the ordered states.
The last equality requires finite $g$ (and $d>1$),
 therefore $b)$ holds only for
systems with the discrete internal symmetry. Finally the
property $c)$ follows directly from the positivity of the
eigenvalues of the transfer matrix since
\begin{equation}
r_L(\beta)= {
              \sum_{\alpha,\beta} \lambda_{\alpha} \lambda_{\beta}
           \over \sum_{\alpha} \lambda_{\alpha}^2
        } > 1  .
\label{pos}
\end{equation}

Characteristic function, Eq.(\ref{rho}), is sensitive to the whole
spectrum of the transfer matrix. However in view of our earlier
discussion supplemented with the properties
$a) - c)$ it is natural to expect that the maximum of $\rho$ occurs
at the phase transition
point,
\begin{equation}
\beta_{max}=\beta_c   .   \label{prop}
\end{equation}
Surprisingly this simple proposition is true in many,
sometimes nontrivial, cases.
We shall discuss them in the order of increasing complexity.

{\em Two dimensional Ising model.}
For $d=2$,  calculation of $\rho(\beta)$ is easily reduced to
solving a straightforward two spin problem.
Indeed the partition functions $Z_i (i=1,2)$ are readily written as
\begin{equation}
Z_i=Tr (T_i)^L ,   \label{smallt}
\end{equation}
where $T_i$ are the transfer matrices propagating one/two spins
{\em horizontally}, c.f. Fig.1.

Explicitly for Ising spins
\begin{eqnarray}
T_1=\left( \begin{array}{cc}
             x^2 & x   \\
             x   & x^2 \\
           \end{array} \right),
T_2=\left(\begin{array}{cccc}
        x^4 & x^2 & x^2 & x^2 \\
        x^2 & x^2 & 1   & x^2 \\
        x^2 & 1   & x^2 & x^2 \\
        x^2 & x^2 & x^2 & x^4 \\
         \end{array}       \right), x=e^\beta.
\end{eqnarray}
We have chosen the interaction energy
\begin{equation}
E(s)=-\sum_{<nm>} \delta_{s_n, s_m}, \label{ip}
\end{equation}
and $s_n=\pm 1$.
In the thermodynamical limit the largest eigenvalues of $T_i$ dominate and
we get
\begin{equation}
\rho_{Ising2}(\beta)=  { t_{1max}^2 \over t_{2max} } = { x^2(x+1)^2
                              \over
\frac{1}{2}(x^2+1)^2+\sqrt{\frac{1}{4}(x^2-1)^4+x^2(x^2+1)^2}} .  \label{i2}
\end{equation}
Since the $L\rightarrow \infty$ limit was already performed, $\rho(\beta)$,
 as given
by Eq.(\ref{i2}), is the nontrivial characteristic of the infinite system.
As conjectured
$\rho(\beta)$ has a single maximum at $x_c=1+\sqrt{2}$ which corresponds
to the famous
Onsager value
\cite{ons,kauf} \footnote{Note the difference by a factor of 2 which
is caused by
our choice of the Potts-like interactions in Eq.(\ref{ip}).} .

{\em Potts and equivalent $Z(N)$ models.}  The energy of the
$q$ state Potts model is given by Eq.(\ref{ip})
with $s_n$ assuming $q$ different values. Transfer matrices for one
and two spin system are again simple
\begin{eqnarray}
<s|T_1|s'>&=&\exp{\beta (\delta_{s,s'}+1)},\\  \nonumber
<s_1,s_2|T_2|s_1',s_2'>&=&
\exp{\beta(\delta_{s_1,s_2}+\delta_{s_1,s_1'}+\delta_{s_2,s_2'}
+\delta_{s_1',s_2'})} . \label{Tpotts}
\end{eqnarray}
For $q=3$ the diagonalization is tractable \footnote{All algebraic
calculations were done
with the aid of MATHEMATICA.} and the final result for the characteristic
function
reads
\begin{eqnarray*}
\lefteqn{\rho_{Potts3}(\beta)=} \\
 & & {2x^2(x+2)^2 \over
x^4+3x^2+2x+3+\sqrt{x^8+2x^6-4x^5+27x^4+28x^3+6x^2+12x+9}  }
\end{eqnarray*}
Again it has a single maximum at $x_c=1+\sqrt{3}$ which agrees with the
known location
of the transition temperature \cite{wu}.

For $q=4,5$
we have used standard numerical methods
to diagonalize transfer matrices. In both cases
$\rho(x)$ has a sigle maximum located at $x_c=1+\sqrt{q}$ in accord with
known results.
For $q > 4$ the transition is  first order \cite{wu} and consequently
our method seems
to apply to both kinds of transitions. We have not attempted algebraic
diagonalization
of $T_i$ for arbitrary $q$.
Low $N$ $(N<5)$ $Z(N)$ (clock) models are equivalent to Potts3 (N=3) and
Ising (N=4)
 systems. Not surprisingly the maximum of $\rho(\beta)$ again agrees
with known results.
We obtain $x^{Z(3)}_c=(1+\sqrt{3})^{2\over 3}$ and $x^{Z(4)}_c=1+\sqrt{2}$
within the accuracy of our numerical procedures.

All previously considered systems were selfdual.
Therefore one may justifiably wonder if our principle is not yet
another manifestation
of the selfduality. Next example demonatrates that the ``maximum rule'',
Eq.(\ref{prop}), is at least more general than the simple duality.

{\em Ising model on a triangular lattice.}
Transfer matrices $T_{1,2}$ \footnote{From now on we use the standard
Ising energy $E(s)=-\sum_{<ij>} s_i s_j$.}
(see Fig.2.)
\begin{eqnarray}
<s_1|T_1|s_1'>&=&\exp{\beta(2 s_1 s_1'+1)},\\  \nonumber
<s_1,s_2|T_2|s_1',s_2'>&=&
\exp{\beta(s_1 s_1'+s_2 s_2'+s_1 s_2 +s_1's_2'+s_1 s_2'+s_2 s_1')}, \nonumber
\end{eqnarray}
can be simply diagonalized. We get
\begin{eqnarray}
t_1^{max}&=&x^3+x^{-1}, \\
t_2^{max}&=&{1\over 2 x^2}(3+x^8+\sqrt{x^{16}-2x^8+16 x^4+1}).
\end{eqnarray}
The maximum of the characteristic function is located at $x_{max}=
3^{1\over4}$.
This agrees with the critical temperature
first derived for this system by Onsager \cite{ons}.
One should remember however that Ising models on triangular and hexagonal
lattices
are interrelated via the duality and star-trangle relations,
hence effectively there exists a symmetry which determines transition
points in both systems \cite{bax}. Our maximum rule could in principle
be a consequence of such a symmetry in this case.
Next application provides more stringent test of this
possibility.

{\em Two layer Ising model.} This system consists of the two planes of
Ising spins coupled
by the nearest neighbour ferromagnetic interaction, also along the vertical
(between the
planes) direction. Periodic boundary conditions  are assumed in all three
directions, which
amounts to doubling the strengh of interaction between the planes.
The system in not selfdual and to our knowledge no other more complicated
symmetries are known.
 Consequently its critical temperature was never derived.  On the contrary
our method provides relatively simple analytic predictions for $\beta_{max}$.
We proceed analogously to the previous cases. Transfer matrices $T_{1,2}$
propagate now states of two and four spins respectively (see Fig.3).
$T_1$ reads
\begin{equation}
<s_1s_2|T_1|s_1's_2'>=\exp{\beta(s_1s_2+s_1s_1'+s_2s_2'+s_1's_2'+2)},
\;\;\;s_i=\pm 1,
\end{equation}
and its largest eigenvalue is
\begin{equation}
t_1^{max}={x^2\over 2}(x^4+2+x^{-4}+\sqrt{x^8+14+x^{-8}}). \label{t1}
\end{equation}
Matrix elements of $T_2$ are
\begin{eqnarray*}
<s_1s_2s_3s_4|T_2|s_1's_2's_3's_4'>=\exp{\beta(s_1s_2+s_2s_3+
s_3s_4+s_4s_1)}\\
\exp{\beta(s_1s_1'+s_2s_2'+s_3s_3'+s_4s_4')}
\exp{\beta(s_1's_2'+s_2's_3'+s_3's_4'+s_4's_1')}.
\end{eqnarray*}
Diagonalisation of this $16\times16$ matrix is simplified noting that
$T_2$ conserves
the $U$-parity, $[T_2,U]=0, U=\prod_{i=1}^4 \sigma_i^x$.
The largest eigenvalue
belongs to the $U=+1$ sector. Final expression is little more complicated
\begin{eqnarray}
t_2^{max}&=&{(1+x^4)^2 \over 4 x^{12}} q_3  +  {1+x^8 \over
4x^{12}}\sqrt{q_1}
+{1+x^4 \over 2\sqrt{2} x^{12}} \sqrt{q_2+(1+x^8) q_3 \sqrt{q_1} }, \\
\label{dlay}
q_1(x)&=&x^{32}-4x^{24}+70x^{16}-4x^8+1,\\ \nonumber
q_2(x)&=&x^{40}-2x^{36}+5x^{32}+26x^{24}+4x^{20}+26x^{16}+5x^8-2x^4+1,\\
\nonumber
q_3(x)&=&x^{16}-2x^{12}+6x^8-2x^4+1. \nonumber
\end{eqnarray}
Resulting characteristic function $\rho(\beta)$ is shown in Fig.4.
It has the single maximum
located at
\begin{equation}
\beta_{max}=0.2656...\;\; .   \label{betac}
\end{equation}
 According to our proposition, Eq.(\ref{prop}), this gives the
transition temperature of the two layer
Ising system.
For comparison:
$\beta_c^{Ising2}\simeq 0.4407...$ and
$\beta_c^{Ising3}\simeq 0.221652(3)$ \cite{has}.
Confronting this number
with the results from MC simulations
would provide the crucial test of our hypothesis. It would be also very
interesting to search
for the "generalized selfduality" - the invariance which would
assure existence of a single maximum at the transition point.
An attractive possibility is to use the characteristic function
given by Eqs.(\ref{t1},\ref{dlay})
 to {\em define } such a mapping. This assumption has
many verifiable consequences. For example, all higher normalized moments of
${\cal T}$ should respect the same symmetry.

{\em Three dimensional Ising model.}  According to our proposition the
transition temperature of
the $d$ dimensional system is determined by the $\beta$ dependence
of the free energies $\beta F_{1,2}=-\log{Z_{1,2}}$ of the corresponding
$d-1$
dimensional
systems. In particular, the problem of finding $\beta_c$ for
the three dimensional Ising
model would be reduced to finding the free energy of the two coupled
layers of
the Ising spins. Indeed in this case the full transfer matrix ${\cal T}$
propagates the whole plane of spins, say, vertically, while the reduced
transfer
 matrices $T_{1,2}$ propagate one (two) rows
horizontally. No analytic solution of this system exists up to date
\footnote{In our previous application we have
derived the transition temperature only.}. However exact expressions
for the complete partition functions $Z_{1,2}(\beta,L)$
{\em in the finite volume} are available for not so large $L$
\cite{bha,stod}.
We have therefore calculated the exact locations
$\beta_{max}(L)$ of the ratio $r_L(\beta)$, c.f. Eq. (\ref{rl}),
which define the pseudocritical temperature
at finite volume, c.f. Table I. \newline
\begin{table}
\begin{tabular}{c|ccccc} \hline\hline
 $L$         &     3  &     4  &     5  &     6  &  7     \\  \hline\hline
 $\beta_{max}(L)$& 0.3317 & 0.3067 & 0.2938 & 0.2859 & 0.2698 \\ \hline\hline
\end{tabular} \newline
\caption{Table I. Volume dependence of the pseudocritical temperature
for the three dimensional Ising model.}
\end{table}
In the thermodynamical limit $\beta_{max}(L)$ should converge to the
true transition
temperature. Even though the available range of $L$ values is rather
limited one sees
the proper trend in the $L$ dependence.  Our values definitely move toward
$\beta_c^{Ising3}$ which was quoted above. One needs larger sizes to
test quantitatively the $L$ dependence against the finite size
scaling predictions.

Variety of approximate methods (Monte Carlo, high temperature expansion)
can  be  also employed to test predictions of Eq.(\ref{prop}),
 in this case.

{\em Limits of applicability.} We have also investigated situations where
the maximum rule does not work.
The regularity emerging from this study indicates that the method does
not apply to
 systems with more than two phases.
We have calculated the characteristic function for the variety of models
with the intermediate Kosterlitz-Thouless (KT) \cite{KT} phase.
The maximum always occured inside the KT region. This phenomenon was found
for $Z(N)$ models with N=5-19, and for the icosahedron and dodecahedron
models
in two dimensions. For the O(2) model $\beta_{max}=1.35...$ well above the
known MC estimate
for the transition between the disordered and KT phases $\beta_c=1.1197(5)$
\cite{pin,gup}.
Interpretation of $\beta_{max}$ located inside the KT phase remains an
open and interesting problem.  On the other hand in the two dimensional
O(3) model $\rho(\beta)$
does not have any maximum in accord with the common wisdom about the lack of
a phase transition in this model. Note that for O(2) and O(3) models the
property c) is not
satisfied and indeed calculated $\rho(\infty) > 1$.
Nevertheless the characteristic function
{\em has} a maximum for $O(2)$ while it is monotonic for $O(3)$.
This parallels the difference
of the phase structures of both models.

{\em Summary.} We have found a surprisingly simple criterion for locating
order-disorder
transition. It is exact for selfdual models.
The method allows for the analytic calculation
in variety of more complicated systems. In particular we give the analytic
estimate of the critical
temperature of the two layer Ising system. Monte Carlo check of this
prediction should be the first step towards more
advanced applications.

Our proposition reduces determination of the critical temperature
of the three dimensional Ising model  to finding
the $\beta$ dependence of the free energy of the two coupled planes
of Ising spins.

 While we are lacking the complete proof of our
hypothesis in general case, many approximated methods can be employed
to test it in
specific applications. High temperature expansion is one
interesting possibility.

One can reformulate the maximum principle in other equivalent ways.
Differentiating
the logarithm of Eq.(\ref{rl}) gives as the condition for the maximum
\begin{equation}
u_2(\beta_c)=u_1(\beta_c),   \label{us}
\end{equation}
where $u_{1,2}$ denotes the density of the internal energy for the
one (two) layer
system.  Analogous relations follow from applying our proposition to
higher moments
of transfer matrix. They all say that at the bulk ($d=3$ say) critical
temperature internal energies
of the interacting and noninteracting planes are equal. According to
our earlier
discussion this statement follows from duality for the two dimensional
Ising and Potts models with planes replaced by the spin chains.
Perhaps the most interesting formulation of Eq.(\ref{us}) for higher moments
of the transfer matrix results in the limit of infinite number of planes.
Then we recover the original two dimensional system and our criterion reads
\begin{equation}
u_{d}(\beta_c^{(d)})=u_{d-1}(\beta_c^{(d)}) \label{uinf},
\end{equation}
where $u_{d}$ denotes the internal energy density of the $d$
dimensional system, and $\beta_c^{(d)}$ corresponds
to its crtitical temperature. As emphasized before, this statement follows
from selfduality in the case of the two dimensional Ising model.
It can be also
checked directly. Indeed
\begin{equation}
u_{2}(\beta_c)=u^{Onsager}_{|2\beta=\log{(1+\sqrt{2})}}=-\sqrt{2}=
u_{1}(\beta_c),
\end{equation}
where
$u^{Onsager}(\beta)=-ctgh(2\beta)\left[1+
{2\kappa^{'}\over \pi} K_1(\kappa)\right]$, $\kappa'=2\tanh{(2\beta)}^2-1$,
$\kappa^2+\kappa'^2=1$, $K_1$ is the elliptic function of the
first kind \cite{huang}, and $u_{1}(\beta)=-\tanh{(\beta)}-1$.
For $d>2$ Eq.(\ref{uinf}) remains unproven similarly
to Eq.(\ref{prop}).

To conclude, there are many unanswered questions and much more work is to be
done, but we feel that it is certainly worth
to undertake this effort.
\vspace*{1cm}

This work was triggered by L. Stodolsky interest in the feasibility
of calculating low moments of the transfer matrix. I would like to thank him
for numerous and stimulating discussions. I also thank E. Seiler and
A. Sokal for instructive discussions.

\section*{Figure Captions}
{\bf Fig.1.} Construction of the transfer matrices ${\cal T}$ and $T_2$
for the two dimensional Ising model. Periodic boundary conditions
are understood. \newline
{\bf Fig.2.} Same as Fig.2 but for the triangular lattice. Bonds
impiled by the periodic boundary conditions are shown explicitly.
\newline
{\bf Fig.3.} Same as Fig.3 but for the two layer Ising model.
\newline
{\bf Fig.4.} Characteristic function for the two layer Ising model.
\newpage
\begin{figure}
\begin{picture}(100,270)(-50,-30)
\multiput(20,50)(36,0){4}{\circle*{5}}
\multiput(20,86)(36,0){4}{\circle*{5}}
\multiput(145,68)(6,0){3}{\circle*{2}}
\multiput(170,50)(0,36){2}{\circle*{5}}
\put (20,50) {\line(124,0){124} }
\put (20,86) {\line(124,0){124} }
\put (50,120) { $ T_2$  }
\put(80,125){\vector(1,0){40}}
\multiput(20,50)(36,0){4}{\line(0,36){36} }
\put(170,50) {\line(0,36){36} }
\put(210,45) {${\cal T}$}
\put(214,60){\vector(0,1){20}}
\put(128,-100){{\Large Fig.1}}
\end{picture}
\end{figure}
\pagebreak
\newpage
\begin{figure}
\begin{picture}(100,270)(-50,-30)
\multiput(20,50)(36,0){4}{\circle*{5}}
\multiput(20,86)(36,0){4}{\circle*{5}}
\multiput(20,122)(36,0){4}{\circle{5}}
\multiput(145,68)(6,0){3}{\circle*{2}}
\multiput(170,50)(0,36){2}{\circle*{5}}
\multiput(206,50)(0,36){3}{\circle{5}}
\put(170,122){\circle{5}}
\multiput(170,50)(0,36){2}{\line(1,0){36}}
\multiput(170,50)(0,36){2}{\line(1,1){36}}
\put (20,50) {\line(124,0){124} }
\put (20,86) {\line(124,0){124} }
\put (50,156) { $ T_2$  }
\put(80,161){\vector(1,0){40}}
\multiput(20,50)(36,0){4}{\line(0,36){72} }
\multiput(20,86)(36,0){3}{\line(1,1){36} }
\multiput(20,50)(36,0){3}{\line(1,1){36} }
\put(128,50) {\line(1,1){16} }
\put(128,86) {\line(1,1){16} }
\put(170,50){\line(0,36){72} }
\put(246,45) {${\cal T}$}
\put(250,60){\vector(0,1){20}}
\put(128,-100){{\Large Fig.2}}
\end{picture}
\end{figure}
\pagebreak
\newpage
\begin{figure}
\begin{picture}(100,270)(-50,-30)
\multiput(20,50)(36,0){4}{\circle*{5}}
\multiput(20,86)(36,0){4}{\circle*{5}}
\multiput(32,68)(36,0){4}{\circle*{5}}
\multiput(32,104)(36,0){4}{\circle*{5}}
\multiput(150,76)(6,0) {3}{\circle*{2}}
\multiput(170,50)(0,36){2}{\circle*{5}}
\multiput(182,68)(0,36){2}{\circle*{5}}
\put (20,50) {\line(124,0){124} }
\put (20,86) {\line(124,0){124} }
\multiput(32,68)(0,36){2}{\line(1,0){124}}
\put (50,138) { $ T_2$  }
\put(80,143){\vector(1,0){40}}
\multiput(20,50)(36,0){4}{\line(0,1){36}}
\multiput(20,50)(36,0){4}{\line(2,3){12}}
\multiput(20,86)(36,0){4}{\line(2,3){12}}
\multiput(32,68)(36,0){4}{\line(0,1){36}}
\multiput(170,50)(12,18){2}{\line(0,1){36}}
\multiput(170,50)(0,36){2}{\line(2,3){12}}
\put(210,45) {${\cal T}$}
\put(214,60){\vector(0,1){20}}
\put(128,-100){{\bf {\Large Fig.3}}}
\end{picture}
\end{figure}
\end{document}